\documentclass[aps,prl,twocolumn,superscriptaddress,showpacs,floatfix,nofootinbib]{revtex4-2}

\usepackage{graphicx}
\usepackage{amsmath,amssymb}
\usepackage{bm}
\usepackage{xcolor}
\usepackage[colorlinks=true,linkcolor=blue!60!black,citecolor=blue!60!black,urlcolor=blue!60!black]{hyperref}
\usepackage{footmisc}

\usepackage{hyperref} 
\usepackage{orcidlink}

\graphicspath{{figs/}}

\newcommand{\vA}{v_{\rm A}}
\newcommand{\dB}{\delta B_0}
\newcommand{\Lpar}{L_\parallel}
\newcommand{\kpar}{k_\parallel}
\newcommand{\Oi}{\Omega_i}

\begin{document}

\title{Equipartition between Alfv\'enic turbulence and energetic ions as the origin of the~$-5$~common spectrum in the heliosphere}

\author{Stanislav Boldyrev\,\orcidlink{0000-0001-6252-5169}}
\email{boldyrev@wisc.edu}
\affiliation{Department of Physics, University of Wisconsin at Madison, Madison, Wisconsin 53706, USA}
\affiliation{Center for Space Plasma Physics, Space Science Institute, Boulder, Colorado 80301, USA}
\author{Daniel Humphrey\,\orcidlink{0009-0002-2889-1493}}
\affiliation{Department of Physics, University of Wisconsin at Madison, Madison, Wisconsin 53706, USA}
\author{Vadim Roytershteyn\,\orcidlink{0000-0003-1745-7587}}
\affiliation{Los Alamos National Laboratory, Los Alamos, New Mexico 87545, USA}

\date{\today}

\begin{abstract}
Suprathermal ion tails throughout the heliosphere follow a ``common spectrum,'' $f(v)\propto v^{-5}$. We show that scale-by-scale equipartition between strong Alfv\'enic turbulence and the ions it accelerates predicts both this exponent and the absolute amplitude, $F(v)=B^2/(16\pi^3 m_i v^5)$, with no free parameters. The prediction agrees within a factor of a few with ACE and Voyager~1 spectra across three decades in magnetic energy. We argue that this saturated level is reached where strong turbulence accelerates particles locally; quiet-time tails lie well below it, consistent with nonlocal acceleration.
\end{abstract}

\maketitle

\paragraph*{Introduction.}
Space and astrophysical plasmas are rarely collisional on the time scales of their dynamics, and their velocity distributions typically consist of a quasi-thermal core and a nonthermal power-law tail~\cite[e.g.,][]{Marsch2006,Oka2018,Verscharen2019}. Such tails are found in Galactic cosmic rays~\cite{BeckerTjus2020}, pulsar wind nebulae~\cite{AtoyanAharonian1996}, relativistic reconnection and turbulence~\cite{SironiSpitkovsky2014,Zhdankin2017,ComissoSironi2018,Vega2022}, black-hole coronae~\cite{Mbarek2026}, and the solar wind, where they are usually described by kappa distributions~\cite{PierrardLazar2010,LivadiotisMcComas2013}. What determines their shape and, more stringently, their intensity is a central question of plasma astrophysics.

The question is also of practical importance. Suprathermal ions are closely linked to the solar energetic particles accelerated at interplanetary shocks and coronal mass ejections, the primary radiation hazard of space weather~\cite{DesaiGiacalone2016}. In the outer heliosphere, the anomalous cosmic rays (ACRs) carry about half of the particle pressure of the heliosheath~\cite{FiskGloeckler2022}. Nonthermal tails also arise in laboratory plasmas~\cite{Magee2019,Hartouni2022,Cruz2018}.

Over the past two decades, measurements with the Solar Wind Ion Composition Spectrometer (SWICS)~\cite{Gloeckler1992} on ACE and Ulysses, and with the Low-Energy Charged Particle (LECP) instruments on Voyager~1 and~2, have uncovered a striking regularity in the distributions of accelerated particles. In cases where the tail is well resolved in observations — such as at low energies downstream of interplanetary shocks, in corotating interaction regions, in the quiet slow and fast solar wind, and throughout the heliosheath up to tens of MeV— it takes a consistent form: a phase-space density $f(v)\propto v^{-5}$, usually terminated by an exponential rollover~\cite{FiskGloeckler2006,Gloeckler2008,MasonGloeckler2012,FiskGloeckler2014,Kollmann2019,FiskGloeckler2022}. Fisk and Gloeckler named this the ``common spectrum.'' Its exponent does not depend on shock strength or heliocentric distance, and it cannot be obtained from diffusive shock acceleration where index is set by the compression ratio, or from conventional stochastic acceleration~\cite{FiskGloeckler2014}. 

A number of explanations have been proposed for the ``common spectrum:'' a thermodynamic ``pump'' mechanism in compressive turbulence~\cite{FiskGloeckler2006,FiskGloeckler2008,FiskGloeckler2014}, with its critique~\cite{JokipiiLee2010}; back-reaction of the accelerated particles on the turbulence~\cite{Bykov2001,ZhangLee2013,Lemoine2023,Mbarek2026}; superposition of stochastic processes~\cite{Schwadron2010}; acceleration in contracting magnetic islands~\cite{Drake2013}; and collisionless relaxation to Lynden-Bell equilibria~\cite{LyndenBell1967,Ewart2023,Ewart2025,Banik2024}. All of these address the power-law spectral exponent. None, to our knowledge, has been confronted with the \emph{absolute} amplitude of in situ spectra.

In this Letter we test a recent hypothesis of Ref.~\cite{Boldyrev2026}, in which the power-law tail is the signature of scale-by-scale energy balance between strong Alfv\'enic turbulence and the particles it accelerates. In its simplest, ``equipartition'' form, the model yields a spectrum whose amplitude depends on the magnetic field alone. It can therefore be compared with observations with no free parameters. In what follows, we first present the physical picture of the equipartition model, and then carry out this comparison in two very different environments: the dense, strongly turbulent plasma downstream of interplanetary shocks at 1~AU and the tenuous, weakly magnetized heliosheath at 120~AU, covering three decades in magnetic energy and four decades in particle energy. We find that the model agrees with the measured spectra in slope and, within a factor of a few, in amplitude. \\

\paragraph*{Equipartition between eddies and particles.}
Consider a nonrelativistic plasma threaded by a mean magnetic field $B_0$ and stirred by strong Alfv\'enic turbulence with outer scale $\Lpar$ along the field and amplitude $\dB$ at that scale. Two properties of such turbulence are essential for our analysis~\cite[e.g.,][]{GoldreichSridhar1995,Boldyrev2006,Chandran2015}. First, the magnetic energy is distributed over field-parallel wave numbers as
\begin{equation}
  E_B(\kpar)\,d\kpar = A_B\,\kpar^{-2}\,d\kpar ,
  \label{eq:spectrum}
\end{equation} 
where $A_B={\dB^2}/{(8\pi\Lpar)}$. Second, the eddies are anisotropic: an eddy of parallel size $l$ is elongated along the local mean magnetic field, its field lines deviating from this field by a small angle $\theta$ obeying the critical-balance relation $\theta^2\approx(\dB/B_0)^2\,l/\Lpar$, where $l=2\pi/\kpar$.

Now place in this turbulence a population of energetic ions with speed $v\gg\vA$, mass $m_i$, and gyrofrequency $\Oi$. What eddies can a particle of given speed interact with most effectively? A particle streaming along a wandering field line responds to an eddy coherently if it traverses the eddy in about one gyroperiod; the smallest eddy that can guide the particle, and hence exchange energy with it through the curvature drift~\cite{Lemoine2023,Vega2024,Boldyrev2026}, is the one in cyclotron resonance with it,
\begin{equation}
  \kpar v = \pm\Oi, \qquad\text{i.e.,}\qquad l=\frac{2\pi v}{\Oi}.
  \label{eq:resonance}
\end{equation}
Faster particles interact with longer, more energetic eddies. Moreover, the particle must move nearly along the local field line to stay inside the thin, elongated eddy, so its pitch angle is of the order of the eddy's own anisotropy angle, 
\begin{equation}
  \theta^2 \approx \frac{\dB^2}{B_0^2}\frac{l}{\Lpar}
             = \frac{\dB^2}{B_0^2}\,\frac{2\pi v}{\Oi\Lpar}.
  \label{eq:angle}
\end{equation}
Only the fraction $\Delta\Omega/4\pi\approx\theta^2/2$ of the particles of speed $v$ (those in the two narrow cones around $\pm\bm B_0$) are thus strongly coupled to the eddy of scale $l(v)$.


The equipartition hypothesis of Ref.~\cite{Boldyrev2026} is now simple to formulate. Suppose that the accelerated ions have grown so energetic that they are no longer passive test particles: the turbulence at scale $l$ cannot deliver to the resonant ions more energy than it contains, and in a turbulent cascade the two reservoirs tend toward equipartition, \emph{scale by scale}. With $n$ the ion density and $f(p)\,dp$ the normalized distribution of ion momenta, the energy density of the resonant ions in the cone is $n\,(p^2/2m_i)\,(\theta^2/2)\,f(p)\,dp$, and the balance reads
\begin{equation}
  A_B\,\kpar^{-2}\,d\kpar \;\approx\; \frac{n p^2}{2m_i}\,\frac{\dB^2}{B_0^2}\,\frac{\pi v}{\Oi\Lpar}\,f(p)\,dp .
  \label{eq:balance}
\end{equation}
Using $\kpar^{-2}d\kpar=-dp/(\Oi m_i)$ from Eq.~\eqref{eq:resonance}, the turbulence amplitude $\dB^2$ and outer scale $\Lpar$ cancel identically, and one obtains the limiting distribution (Eq.~(12) of Ref.~\cite{Boldyrev2026}):
\begin{equation}
  f(p)\,dp \approx \frac{1}{4\pi^2}\,\frac{m_i^2\vA^2}{p^5}\;4\pi p^2\,dp ,
  \label{eq:fp}
\end{equation}
where $\vA={B_0}/{\sqrt{4\pi n m_i}}$ is the Alfv\'en speed. Expressed as a phase-space density in velocity space, $n f(p)\,dp = F(v)\,4\pi v^2 dv$, this is
\begin{equation}
F(v)= \frac{n_0\,\vA^2}{4\pi^2\,v^5}=\frac{B_0^2}{16\pi^3 m_i\,v^5},
  \label{eq:Fv}
\end{equation}
with $n_0$ the ion density. 

Equation~\eqref{eq:Fv} is the central result we are going to test. Two remarks are in order here. First, Eq.~(\ref{eq:Fv}) is 
 a phenomenological estimate, and its numerical coefficient is fixed only up to a factor of order unity. Agreement with observations to within such a factor is therefore considered consistent with the model.  Second, and more unusually, the turbulence level itself, $\dB^2$, drops out, and the amplitude is fixed by the magnetic field alone as the density cancels through $n_0\vA^2=B_0^2/4\pi m_i$.  In the conventional terms used by energetic-particle instruments, this relationship translates to the differential intensity per unit kinetic energy $T$, as follows: 
\begin{eqnarray}
j(T)=\frac{v^2}{m_i}F(v)
      =\frac{B_0^2}{32\sqrt{2}\,\pi^3\,m_i^{1/2}}\,T^{-3/2}, 
\label{eq:jT}
\end{eqnarray}
which results in an intensity spectrum with a slope of~$-1.5$.
\\
\begin{figure*}[t]
\centering
\includegraphics[width=0.9\textwidth]{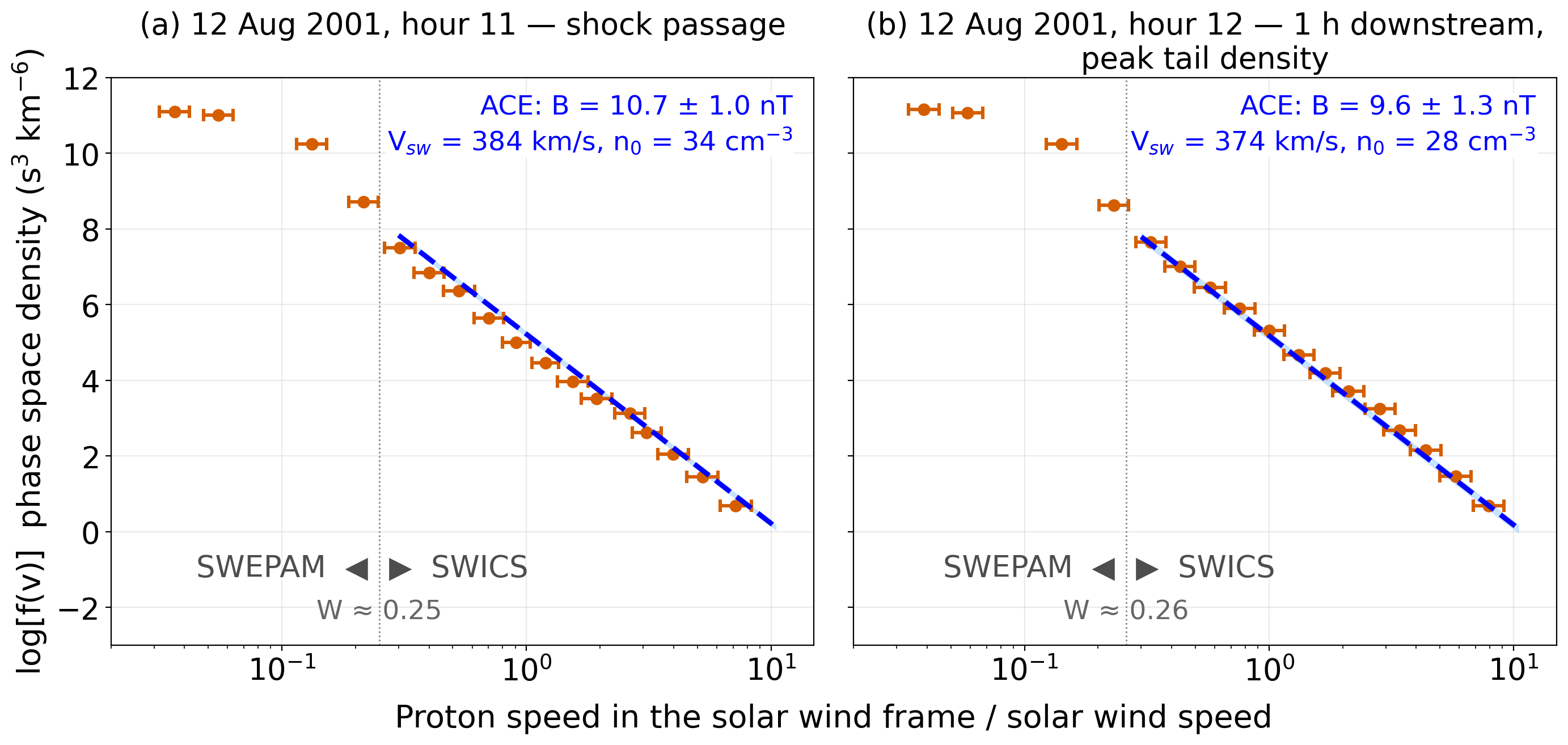}
\caption{Solar-wind-frame proton phase-space density measured by ACE/SWICS and SWEPAM during the 12 August 2001 local acceleration event, (left) in the hour of shock passage and (right) one hour downstream, at the peak of the tail density [data points digitized from Fig. 3 of Ref.~\cite{FiskGloeckler2014} and replotted.]  The blue dashed line is the equipartition prediction, Eq.~\eqref{eq:Fv}, evaluated with the measured $B=10.7$~nT and $V_{\rm sw} = 384$~km\,s$^{-1}$ used for the abscissa~(left), and $B=9.6$~nT and $V_{\rm sw} = 374$~km\,s$^{-1}$ used for the abscissa~(right). The shaded band shows the range of the prediction for $B$ varied by $\pm 1\sigma$ of the 1-min field magnitude within each hour ($\pm 1.0$ and $\pm 1.3$ nT, respectively).}
\label{fig:aug12}
\end{figure*}

\paragraph*{Agreement at 1~AU: the prime acceleration regions.}
Fisk and Gloeckler~\cite{FiskGloeckler2014} identified the observations that most directly expose the acceleration process: proton spectra measured by ACE/SWICS within the ``prime acceleration regions'' (PARs) of local acceleration events, i.e., the hot spots of enhanced density and magnetic turbulence that follow strong interplanetary shocks. In these regions, the tail intensity peaks and contamination by remotely accelerated particles is negligible. There, the tail is well approximated by a power law extending from the bulk of solar wind to the end of the SWICS energy range, and the measured spectral index is close to~$-5$.

Figure~\ref{fig:aug12} shows the 12 August 2001 event, redrawn from \citet[][their Fig.~3]{FiskGloeckler2014}. The blue dashed line shows the prediction of our Eq.~\eqref{eq:Fv}, which contains no free parameters once the plasma conditions are specified; it is set by the magnetic field strength alone, with the solar wind speed entering only through the conversion $w = v/V_{\rm sw}$. As the inputs to the model, we use the hourly means of ACE MAG and SWEPAM data for the same intervals:\footnote{To evaluate the parameters for the ACE spectra, we used the Level-2 data of the ACE magnetometer (MAG, 16-s resolution; AC\_H0\_MFI, N. Ness/C. W. Smith) and of the Solar Wind Electron, Proton and Alpha Monitor (SWEPAM, 64-s resolution; AC\_H0\_SWE, D. J. McComas), obtained from NASA’s Coordinated Data Analysis Web (CDAWeb, https://cdaweb.gsfc.nasa.gov).\label{footnote_1}} (a)~$B = 10.7\,{\rm nT}$, $n_0 = 34\,{\rm cm^{-3}}$, $V_{\rm sw} = 384 \,{\rm km\, s^{-1}}$, giving $v_A = 40\,{\rm km\, s^{-1}}$, and (b)~$B = 9.6\, {\rm nT}$, $n_0 = 28\, {\rm cm^{-3}}$, $V_{\rm sw} = 374 \,{\rm km\,s^{-1}}$, giving $v_A = 39\, {\rm km\,s^{-1}}$. The agreement is very good: averaged over $w > 1$, the observed tail falls to ~$\sim 0.6$ of the prediction in~(a) and exceeds the prediction by a factor~$\sim 1.4$ in~(b).
\begin{figure*}[t]
\centering
\includegraphics[width=0.90\textwidth]{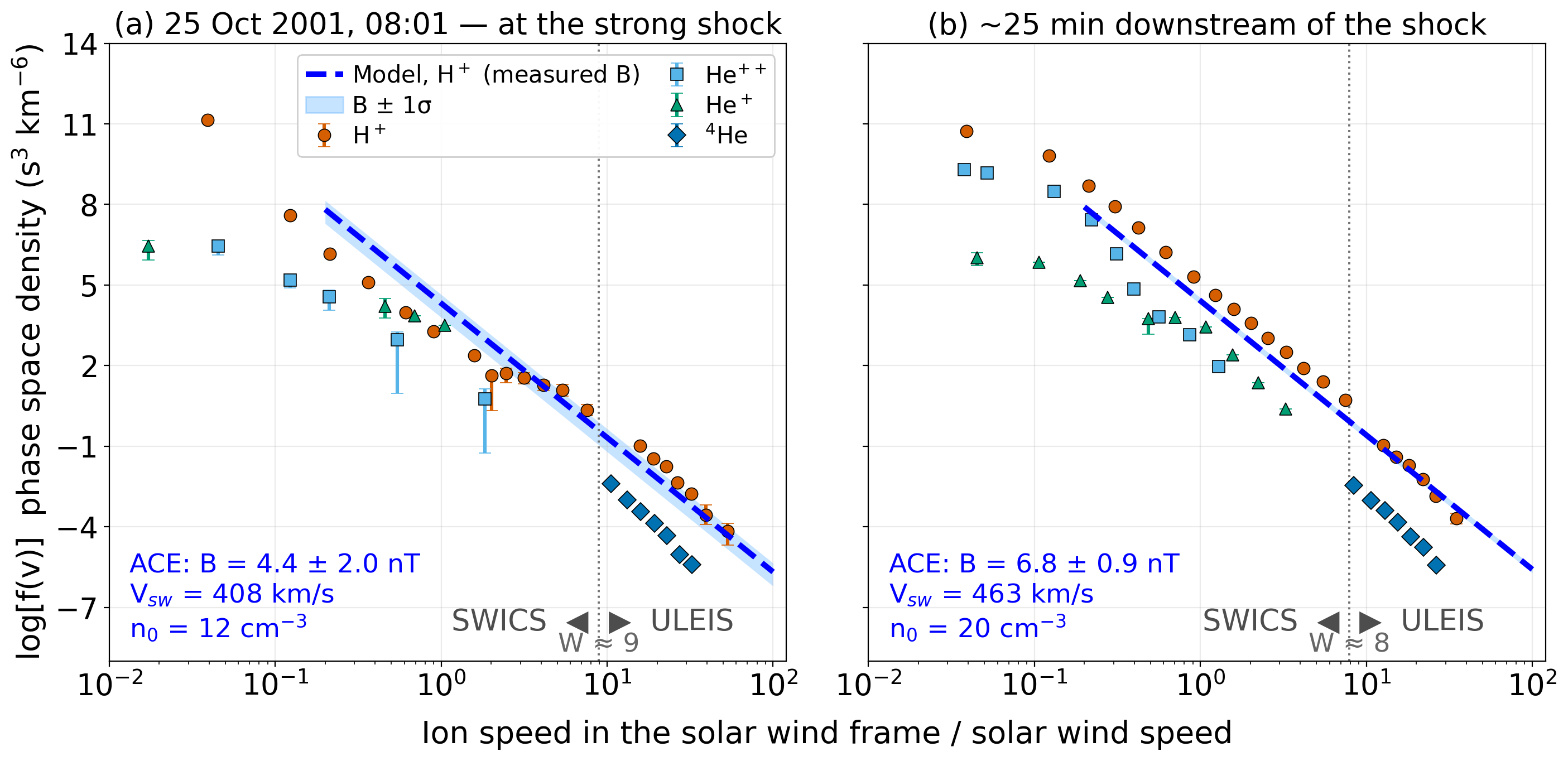}
\caption{Twelve-minute, sector-averaged solar-wind-frame distribution functions of H$^+$, He$^{++}$, He$^+$, and $^4$He from ACE/SWICS and ULEIS during the 25 October 2001 local acceleration event, (left) at shock passage and (right) $\sim25$~min downstream, inside the first prime acceleration region [data points digitized from Fig. 5 of Ref.~\cite{FiskGloeckler2014} and replotted.]
The blue dashed line is the prediction of Eq.~(\ref{eq:Fv}) for protons. The inputs are evaluated with:
(a)~$B=4.4$~nT, $V_{\rm sw}=408$~km\,s$^{-1}$;
(b)~$B=6.8$~nT, $V_{\rm sw}=463$~km\,s$^{-1}$.
The shaded bands correspond to $B\pm1\sigma$ of the 1-min field magnitude within each interval ($\pm 2.0$ and $\pm0.9$~nT, respectively).}
\label{fig:oct25}
\end{figure*}

Figure~\ref{fig:oct25} shows the 25 October 2001 event, associated with a strong quasi-parallel shock, for which the SWICS and ULEIS measurements together span more than ten decades in phase-space density, redrawn from \citet[][their Fig. 5]{FiskGloeckler2014}: (a) the 12-min interval of shock passage (08:01 UT) and (b) a 12-min interval ~25 min downstream of the shock, in the first prime acceleration region. The blue dashed line is the prediction of our Eq.~(\ref{eq:Fv}) for protons. The inputs\footref{footnote_1} are the means of the
ACE MAG and SWEPAM 1-min binned data over each spectrum's interval:
(a)~$B=4.4$~nT, $n_0=12\,{\rm cm^{-3}}$, $V_{\rm sw}=408$~km\,s$^{-1}$,
$\vA=28$~km\,s$^{-1}$ (this window straddles the shock, where $B$ rises from
2.5 to 7.5~nT, hence the large scatter);
(b)~$B=6.8$~nT, $n_0=20\,{\rm cm^{-3}}$, $V_{\rm sw}=463$~km\,s$^{-1}$,
$\vA=33$~km\,s$^{-1}$. 
The log-log tails are less straight in these measurements; in (a), the observed proton tail lies below the power-law prediction by a factor of $\sim 2$, reflecting the modulated spectrum
at the shock itself; in (b), it lies above by a factor of~$\sim 4.5$, averaged over $w>1$.
\\
\begin{figure*}[t]
\includegraphics[width=0.9\textwidth]{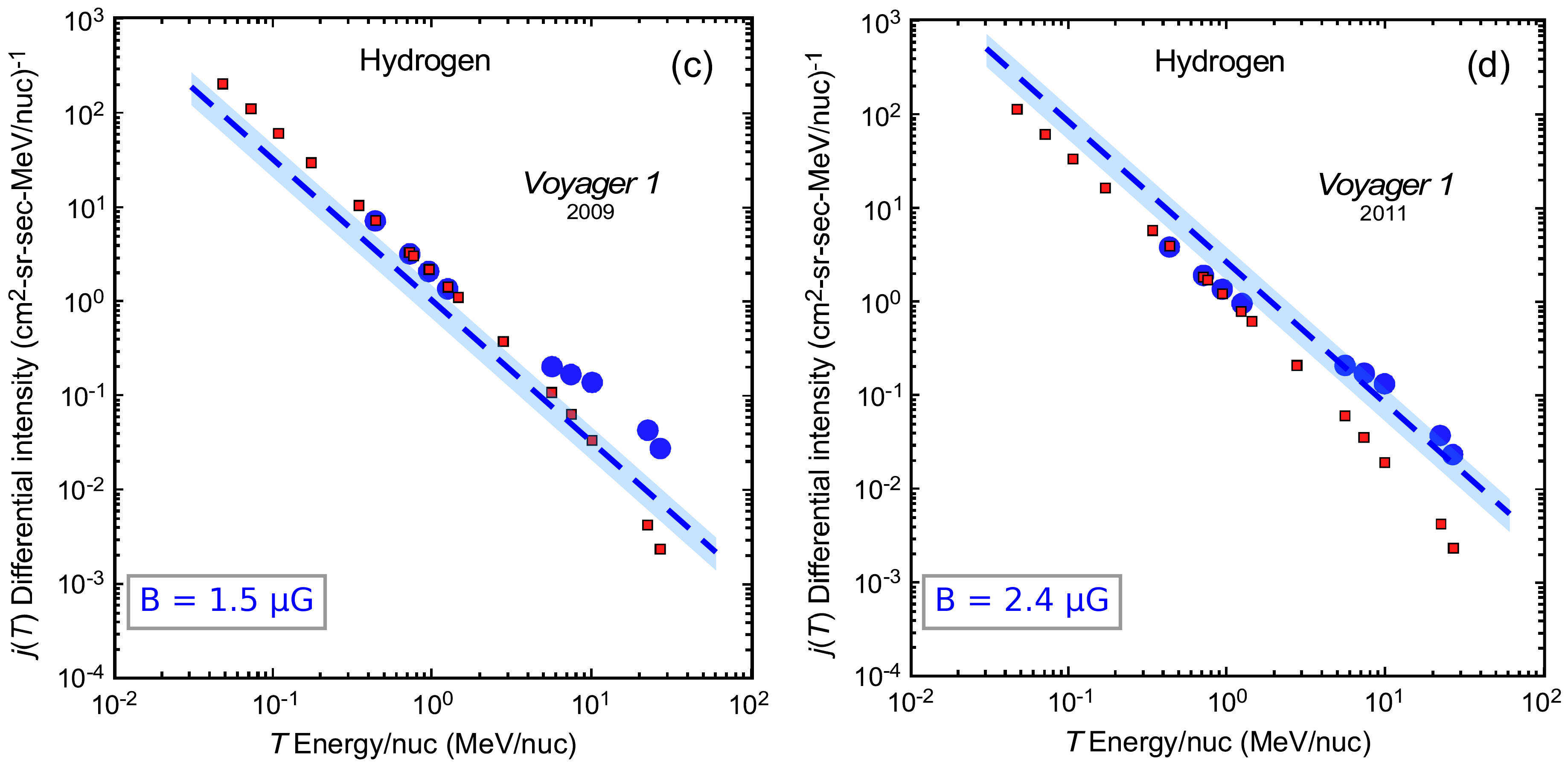}
\caption{Differential intensity of ACR hydrogen measured by Voyager~1/LECP in the heliosheath, averaged over (left) 2009, two-thirds of the way between the termination shock and the ACR boundary, and (right) 2011, the last full year before that boundary [adapted from Fig.~2 of Fisk \& Gloeckler~(2022) ``Global Structure and Dominant Particle Acceleration Mechanism of the Heliosheath:
Definitive Conclusions," Ref.~\cite{FiskGloeckler2022}, published under a CC~BY~4.0 license]. 
Blue points are the LECP hydrogen data above 400~keV; red points are the original authors' $T^{-1.5}$ fits extended to lower energies. The added blue dashed line is the analytical prediction given by Eq.~\eqref{eq:jT} evaluated with the yearly mean magnetic field of Ref.~\cite{FiskGloeckler2022}, $B=1.5$~$\mu$G (2009) and $2.4$~$\mu$G (2011); the band corresponds to $\pm20\%$ in $B$.}
\label{fig:voyager}
\end{figure*}

\paragraph*{Agreement at 120~AU: anomalous cosmic rays in the heliosheath.}
The heliosheath, between the solar-wind termination shock at $\sim 90$~AU and the heliopause, is the largest acceleration region in the heliosphere. There the interstellar pickup ions, heated at the termination shock, are accelerated into the ACRs, whose differential intensity spectrum is a $T^{-1.5}$ power law from below 100~keV to a rollover at $\sim 15$--$100$~MeV~\cite{Decker2005,FiskGloeckler2014,FiskGloeckler2022}. Fisk and Gloeckler~\cite{FiskGloeckler2022} have recently re-analyzed the Voyager~1/LECP hydrogen data and fitted yearly averaged spectra of the form $j(T)=j_0T^{-1.5}\exp(-T/T_{\rm roll})$ at four positions along the trajectory, together with the yearly mean magnetic field from the V1 magnetometer: $B\simeq1$~$\mu$G in 2005 and 2007, $1.5$~$\mu$G in 2009 and $2.4$~$\mu$G in 2011, at 110--120~AU. 
These spectra provide a test of the prediction given by Eq.~\eqref{eq:jT} in a regime removed by three orders of magnitude in~$B^2$ and by four orders of magnitude in particle energy from the PARs at 1~AU. The test is particularly clean because the prediction needs no density: the plasma density in the heliosheath is poorly known, but the magnetic field is measured.

Figure~\ref{fig:voyager} shows the 2009 and 2011 spectra with Eq.~\eqref{eq:jT} overlaid. These measurements are made deep in the inner heliosheath, where the plasma had resided for years, and the flow had nearly stagnated, so these are the measurements most representative of in situ acceleration by local turbulence rather than of shock-processed plasma. With $B=1.5$~$\mu$G the predicted intensity at 1~MeV is $0.98\,{\rm (cm^2\,sr\,s\,MeV)^{-1}}$, against the fitted $j_0=2.1$; with $B=2.4$~$\mu$G it is $2.5$, against $j_0=1.15$. The predicted line thus runs through the measured ACR spectrum, with the correct slope, over the four decades in energy below the rollover, and the absolute level is reproduced to within a factor of two in both years. (The other two spectra analyzed in \cite[][]{FiskGloeckler2022} and not shown here, $B=1$~$\mu$G spectra of 2005 and 2007, with $j_0=2.3$, lie a factor $\simeq5$ above the prediction; the departure grows toward the termination shock, where the just-heated pickup ions, not yet pumped into the tail, carry most of the pressure~\cite{FiskGloeckler2022}.)

Two remarks on the heliosheath comparison are in order here. First, there the particle pressure exceeds the magnetic pressure several-fold [Table~1 of Ref.~\cite{FiskGloeckler2022}], so the equipartition is not between the tail and the bulk magnetic energy but, as in Eq.~\eqref{eq:balance}, between the resonant tail particles and the fluctuating field at the resonant scales; that the tail amplitude nevertheless tracks $B^2$, with no reference to the particle pressure, is a nontrivial check of the mechanism. Second, the year-to-year trend near the ACR boundary is not reproduced by the model: between 2009 and 2011 the field rose while $j_0$ fell, a decline that Fisk and Gloeckler attribute to the escape of ACRs along field lines connecting to the heliopause. The equipartition level given by Eq.~(\ref{eq:jT}) thus provides an upper bound; where particles escape, the tail falls below it.\\

\paragraph*{Quiet-time tails and open questions.}
The common spectrum is also observed far from any shock, for example, in the quiet-time slow solar wind at $4.8$~AU [Fig.~1 of Ref.~\cite{FiskGloeckler2006}] and in the fast polar coronal-hole wind during extended periods selected such that no shocks or local acceleration events were present [Fig.~7 of Ref.~\cite{FiskGloeckler2014}]. A comparison with these spectra (not shown) indicates that Eq.~(\ref{eq:Fv}), evaluated with typical fields for these streams ($B\simeq1$ and $1.4$~nT, respectively), reproduces the $-5$ slope but significantly overestimates the level (observed-to-predicted ratios $\sim5\times10^{-3}$ and $\sim10^{-3}$).



We suggest that the contrast between the two groups of observations may have the following physical explanation. Equation~\eqref{eq:Fv} is not an equilibrium that every plasma must attain; but rather it is an \emph{upper bound}, the saturated state in which the energetic ions get in equipartition with the turbulence at each scale. Reaching it requires that acceleration proceed locally and for long enough time so that the eddy-particle energy exchange becomes strong. This is exactly the situation in the prime acceleration regions of Figs.~\ref{fig:aug12} and~\ref{fig:oct25} (dense, highly compressed, strongly turbulent plasma immediately downstream of a strong shock, where the tail density rises by two orders of magnitude over a few hours) and in the heliosheath of Fig.~\ref{fig:voyager} (a closed, nearly static volume in which the pickup ions are continuously pumped into the tail until the tail holds half the particle pressure). 

The quiet-time tails are a different kind of population. They are accumulated over months to years in the absence of local acceleration events. They may therefore be most naturally interpreted as particles accelerated \emph{elsewhere} --- in distant compressions, corotating interaction regions, or earlier transient events --- that have subsequently spread, by diffusion along and across the field and by adiabatic cooling, into the regions where they are measured. 
A similar point---that regions of strong turbulence are not space-filling, so that volume-averaged spectra are depressed relative to the local saturated one---has been made for kinetic simulations of relativistic turbulence~\cite{Vega2022,Vega2023,Boldyrev2026}.

A second open question concerns composition. In the 25 October 2001 event (Fig.~\ref{fig:oct25}) the He$^{++}$, He$^{+}$, and $^4$He spectra have the same $-5$ slope as the protons but lower amplitudes, in rough proportion to their abundances~\cite{FiskGloeckler2014}.  How the equipartition budget is shared among several species that resonate with the same eddies at different speeds, and whether this reproduces the observed ordering of the heavy-ion tails, is not addressed by the present model and requires further analysis. An evident next step would be to extend the energy-balance argument to a multi-species plasma. Additionally, it would be interesting to explore the relativistic regimes discussed in~\cite[e.g.,][]{Zhdankin2017,Zhdankin2022,ComissoSironi2018,Lemoine2023,Sebastian2025,Boldyrev2026,Mbarek2026}.\\

\paragraph*{Conclusion.}
We have demonstrated that a phenomenological ``equipartition principle''---that strong Alfv\'enic turbulence and the ions it accelerates come to share their energy scale by scale---predicts the suprathermal ion spectrum of the heliosphere in both exponent and absolute level, $F(v)=B_0^2/(16\pi^3 m_i v^5)$, with the magnetic field as its only input. Where particles are accelerated locally by strong turbulence, in the prime acceleration regions downstream of interplanetary shocks at 1~AU and in the heliosheath at 120~AU, the prediction agrees with ACE/SWICS and Voyager/LECP measurements over three decades in magnetic energy and four decades in particle energy, within a factor of a few. To our knowledge, this is the first instance in which the amplitude of the common spectrum, and not only its exponent, has been reproduced, within the order-unity precision of the model, by a theory without adjustable parameters. Our analysis supports the original arguments by Fisk \& Gloeckler that strong turbulent events are the primary, in situ drivers of the ``common spectrum.'' Our model proposes that the long-standing ``common spectrum'' may be viewed as a thermodynamic-like saturation level of the eddy--particle system rather than as a signature of one particular acceleration mechanism. 

The success of the equipartition model has raised important questions that remain unanswered: why the quiet-time tails of the slow and fast solar wind sit well below the saturated level, which we have argued reflects nonlocal acceleration and transport; why the tail near the heliosheath boundary falls as the field rises, which we attribute to escape; and how the energy budget is partitioned among ion species. These questions can be explored using existing data and kinetic simulations. Their more detailed study has the potential to turn the equipartition principle from an empirical success into a predictive tool for nonthermal distributions in space, astrophysical, and laboratory plasmas.\\

\begin{acknowledgments}
We acknowledge the use of ACE/SWICS, ULEIS, and Voyager 1/LECP spectra published in Refs.~\cite{FiskGloeckler2014,FiskGloeckler2022}, and of ACE MAG and SWEPAM data provided by the instrument teams through NASA’s CDAWeb. The observational particle data shown in Figs.~1~and~2 were digitized from the published figures of Ref.~\cite[][]{FiskGloeckler2014} and replotted alongside the theory prediction with the assistance of Claude Opus~5.5 (Anthropic); the authors checked the extracted values against the original figures. The same software was used to overlay the theory prediction with the data of Ref.~\cite{FiskGloeckler2022} in Fig.~\ref{fig:voyager}.  This work was supported by the U.S. Department of Energy, Office of Science, Office of Fusion Energy Sciences under award number DE-SC0024362. The work of D.H. and S.B. was also supported by the University of Wisconsin--Madison, Office of the Vice Chancellor for Research, with funding from the Wisconsin Alumni Research Foundation. 
\end{acknowledgments}

\bibliographystyle{apsrev4-2}
\bibliography{refs}

@ARTICLE{Sebastian2025,
       author = {{Sebastian}, Samuel T. and {Comisso}, Luca},
        title = "{Magnetic Field-line Curvature and Its Role in Particle Acceleration by Magnetically Dominated Turbulence}",
      journal = {Astrophys. J. Lett.},
         year = 2025,
        month = nov,
       volume = {994},
       number = {1},
          eid = {L1},
        pages = {L1},
          doi = {10.3847/2041-8213/ae1696},
archivePrefix = {arXiv},
       eprint = {2510.20628},
 primaryClass = {astro-ph.HE},
       adsurl = {https://ui.adsabs.harvard.edu/abs/2025ApJ...994L...1S}
}

@ARTICLE{Vega2023,
       author = {{Vega}, Cristian and {Boldyrev}, Stanislav and {Roytershteyn}, Vadim},
        title = "{Spatial Intermittency of Particle Distribution in Relativistic Plasma Turbulence}",
      journal = {Astrophys. J.},
         year = 2023,
        month = jun,
       volume = {949},
       number = {2},
          eid = {98},
        pages = {98},
          doi = {10.3847/1538-4357/accd73},
archivePrefix = {arXiv},
       eprint = {2304.11000},
 primaryClass = {physics.plasm-ph},
       adsurl = {https://ui.adsabs.harvard.edu/abs/2023ApJ...949...98V}
}

@ARTICLE{Boldyrev2026,
       author = {{Boldyrev}, Stanislav and {Humphrey}, Daniel and {Roytershteyn}, Vadim and {Vega}, Cristian},
        title = "{Distributions of Particles Accelerated by Strong Alfv{\'e}nic Turbulence}",
      journal = {Astrophys. J. Lett.},
         year = 2026,
        month = aug,
       volume = {1007},
       number = {1},
          eid = {L26},
        pages = {L26},
          doi = {10.3847/2041-8213/ae8bac},
archivePrefix = {arXiv},
       eprint = {2605.02219},
 primaryClass = {physics.plasm-ph},
       adsurl = {https://ui.adsabs.harvard.edu/abs/2026ApJ..1007L..26B}
}

@article{FiskGloeckler2006,
  author  = {Fisk, L. A. and Gloeckler, G.},
  title   = {The common spectrum for accelerated ions in the quiet-time solar wind},
  journal = {Astrophys. J.},
  volume  = {640},
  pages   = {L79--L82},
  year    = {2006},
  doi     = {10.1086/503293}
}

@article{FiskGloeckler2014,
  author  = {Fisk, L. A. and Gloeckler, G.},
  title   = {The case for a common spectrum of particles accelerated in the heliosphere: Observations and theory},
  journal = {J. Geophys. Res. Space Phys.},
  volume  = {119},
  pages   = {8733--8749},
  year    = {2014},
  doi     = {10.1002/2014JA020426}
}

@article{FiskGloeckler2008,
  author  = {Fisk, L. A. and Gloeckler, G.},
  title   = {Acceleration of suprathermal tails in the solar wind},
  journal = {Astrophys. J.},
  volume  = {686},
  pages   = {1466--1473},
  year    = {2008},
  doi     = {10.1086/591543}
}

@inproceedings{Gloeckler2008,
  author    = {Gloeckler, G. and Fisk, L. A. and Mason, G. M. and Hill, M. E.},
  title     = {Formation of power law tail with spectral index $-5$ inside and beyond the heliosphere},
  booktitle = {Particle Acceleration and Transport in the Heliosheath and Beyond},
  series    = {AIP Conf. Proc.},
  volume    = {1039},
  pages     = {367--374},
  publisher = {AIP},
  address   = {Melville, NY},
  year      = {2008},
  doi       = {10.1063/1.2982473}
}

@article{Gloeckler1992,
  author  = {Gloeckler, G. and Geiss, J. and Balsiger, H. and others},
  title   = {The {S}olar {W}ind {I}on {C}omposition {S}pectrometer},
  journal = {Astron. Astrophys. Suppl. Ser.},
  volume  = {92},
  pages   = {267--289},
  year    = {1992}
}

@article{MasonGloeckler2012,
  author  = {Mason, G. M. and Gloeckler, G.},
  title   = {Power law distributions of suprathermal ions in the quiet solar wind},
  journal = {Space Sci. Rev.},
  volume  = {172},
  pages   = {241--251},
  year    = {2012},
  doi     = {10.1007/s11214-010-9741-0}
}

@article{Kollmann2019,
  author  = {Kollmann, P. and Hill, M. E. and McNutt, R. L. and others},
  title   = {Suprathermal ions in the outer heliosphere},
  journal = {Astrophys. J.},
  volume  = {876},
  pages   = {46},
  year    = {2019},
  doi     = {10.3847/1538-4357/ab125f}
}

@article{Ewart2025,
  author  = {Ewart, R. J. and Nastac, M. L. and Bilbao, P. J. and Silva, T. and Silva, L. O. and Schekochihin, A. A.},
  title   = {Relaxation to universal non-{M}axwellian equilibria in a collisionless plasma},
  journal = {Proc. Natl. Acad. Sci. U.S.A.},
  volume  = {122},
  pages   = {e2417813122},
  year    = {2025},
  doi     = {10.1073/pnas.2417813122}
}

@article{Ewart2023,
  author  = {Ewart, R. J. and Nastac, M. L. and Schekochihin, A. A.},
  title   = {Non-thermal particle acceleration and power-law tails via relaxation to universal {L}ynden-{B}ell equilibria},
  journal = {J. Plasma Phys.},
  volume  = {89},
  pages   = {905890516},
  year    = {2023},
  doi     = {10.1017/S0022377823000983}
}

@article{LyndenBell1967,
  author  = {Lynden-Bell, D.},
  title   = {Statistical mechanics of violent relaxation in stellar systems},
  journal = {Mon. Not. R. Astron. Soc.},
  volume  = {136},
  pages   = {101},
  year    = {1967}
}

@article{Bykov2001,
  author  = {Bykov, A. M.},
  title   = {Particle acceleration and nonthermal phenomena in superbubbles},
  journal = {Space Sci. Rev.},
  volume  = {99},
  pages   = {317--326},
  year    = {2001},
  doi     = {10.1023/A:1013817721725}
}

@article{JokipiiLee2010,
  author  = {Jokipii, J. R. and Lee, M. A.},
  title   = {Compression acceleration in astrophysical plasmas and the production of $f(v)\propto v^{-5}$ spectra},
  journal = {Astrophys. J.},
  volume  = {713},
  pages   = {475--483},
  year    = {2010},
  doi     = {10.1088/0004-637X/713/1/475}
}

@article{Drake2013,
  author  = {Drake, J. F. and Swisdak, M. and Fermo, R.},
  title   = {The power-law spectra of energetic particles during multi-island magnetic reconnection},
  journal = {Astrophys. J. Lett.},
  volume  = {763},
  pages   = {L5},
  year    = {2013},
  doi     = {10.1088/2041-8205/763/1/L5}
}

@article{ZhangLee2013,
  author  = {Zhang, M. and Lee, M. A.},
  title   = {Stochastic acceleration of energetic particles in the heliosphere},
  journal = {Space Sci. Rev.},
  volume  = {176},
  pages   = {133--146},
  year    = {2013},
  doi     = {10.1007/s11214-011-9754-3}
}

@article{Schwadron2010,
  author  = {Schwadron, N. A. and Dayeh, M. A. and Desai, M. and Fahr, H. and Jokipii, J. R. and Lee, M. A.},
  title   = {Superposition of stochastic processes and the resulting particle distributions},
  journal = {Astrophys. J.},
  volume  = {713},
  pages   = {1386--1392},
  year    = {2010},
  doi     = {10.1088/0004-637X/713/2/1386}
}

@article{LivadiotisMcComas2013,
  author  = {Livadiotis, G. and McComas, D. J.},
  title   = {Understanding kappa distributions: A toolbox for space science and applications},
  journal = {Space Sci. Rev.},
  volume  = {175},
  pages   = {183--214},
  year    = {2013},
  doi     = {10.1007/s11214-013-9982-9}
}

@article{PierrardLazar2010,
  author  = {Pierrard, V. and Lazar, M.},
  title   = {Kappa distributions: Theory and applications in space plasmas},
  journal = {Sol. Phys.},
  volume  = {267},
  pages   = {153--174},
  year    = {2010},
  doi     = {10.1007/s11207-010-9640-2}
}

@article{Oka2018,
  author  = {Oka, M. and Birn, J. and Battaglia, M. and others},
  title   = {Electron power-law spectra in solar and space plasmas},
  journal = {Space Sci. Rev.},
  volume  = {214},
  pages   = {82},
  year    = {2018},
  doi     = {10.1007/s11214-018-0515-4}
}

@article{Marsch2006,
  author  = {Marsch, E.},
  title   = {Kinetic physics of the solar corona and solar wind},
  journal = {Living Rev. Sol. Phys.},
  volume  = {3},
  pages   = {1},
  year    = {2006}
}

@article{Verscharen2019,
  author  = {Verscharen, D. and Klein, K. G. and Maruca, B. A.},
  title   = {The multi-scale nature of the solar wind},
  journal = {Living Rev. Sol. Phys.},
  volume  = {16},
  pages   = {5},
  year    = {2019},
  doi     = {10.1007/s41116-019-0021-0}
}

@article{BeckerTjus2020,
  author  = {Becker Tjus, J. and Merten, L.},
  title   = {Closing in on the origin of galactic cosmic rays using multimessenger information},
  journal = {Phys. Rep.},
  volume  = {872},
  pages   = {1--98},
  year    = {2020},
  doi     = {10.1016/j.physrep.2020.05.002}
}

@article{AtoyanAharonian1996,
  author  = {Atoyan, A. M. and Aharonian, F. A.},
  title   = {On the mechanisms of gamma radiation in the {C}rab {N}ebula},
  journal = {Mon. Not. R. Astron. Soc.},
  volume  = {278},
  pages   = {525--541},
  year    = {1996},
  doi     = {10.1093/mnras/278.2.525}
}

@article{SironiSpitkovsky2014,
  author  = {Sironi, L. and Spitkovsky, A.},
  title   = {Relativistic reconnection: An efficient source of non-thermal particles},
  journal = {Astrophys. J. Lett.},
  volume  = {783},
  pages   = {L21},
  year    = {2014},
  doi     = {10.1088/2041-8205/783/1/L21}
}

@article{Zhdankin2017,
  author  = {Zhdankin, V. and Werner, G. R. and Uzdensky, D. A. and Begelman, M. C.},
  title   = {Kinetic turbulence in relativistic plasma: From thermal bath to nonthermal continuum},
  journal = {Phys. Rev. Lett.},
  volume  = {118},
  pages   = {055103},
  year    = {2017},
  doi     = {10.1103/PhysRevLett.118.055103}
}

@article{ComissoSironi2018,
  author  = {Comisso, L. and Sironi, L.},
  title   = {Particle acceleration in relativistic plasma turbulence},
  journal = {Phys. Rev. Lett.},
  volume  = {121},
  pages   = {255101},
  year    = {2018},
  doi     = {10.1103/PhysRevLett.121.255101}
}

@article{Vega2022,
  author  = {Vega, C. and Boldyrev, S. and Roytershteyn, V. and Medvedev, M.},
  title   = {Turbulence and particle acceleration in a relativistic plasma},
  journal = {Astrophys. J. Lett.},
  volume  = {924},
  pages   = {L19},
  year    = {2022},
  doi     = {10.3847/2041-8213/ac441e}
}

@article{Vega2024,
  author  = {Vega, C. and Boldyrev, S. and Roytershteyn, V.},
  title   = {Particle acceleration in relativistic {A}lfv\'enic turbulence},
  journal = {Astrophys. J.},
  volume  = {971},
  pages   = {106},
  year    = {2024},
  doi     = {10.3847/1538-4357/ad5f8f}
}

@article{Lemoine2023,
  author  = {Lemoine, M.},
  title   = {Particle acceleration in strong {MHD} turbulence},
  journal = {J. Plasma Phys.},
  volume  = {89},
  pages   = {175890501},
  year    = {2023},
  doi     = {10.1017/S0022377823000946}
}

@article{GoldreichSridhar1995,
  author  = {Goldreich, P. and Sridhar, S.},
  title   = {Toward a theory of interstellar turbulence. {II}. {S}trong {A}lfv\'enic turbulence},
  journal = {Astrophys. J.},
  volume  = {438},
  pages   = {763--775},
  year    = {1995},
  doi     = {10.1086/175121}
}

@article{Boldyrev2006,
  author  = {Boldyrev, S.},
  title   = {Spectrum of magnetohydrodynamic turbulence},
  journal = {Phys. Rev. Lett.},
  volume  = {96},
  pages   = {115002},
  year    = {2006},
  doi     = {10.1103/PhysRevLett.96.115002}
}

@article{Chandran2015,
  author  = {Chandran, B. D. G. and Schekochihin, A. A. and Mallet, A.},
  title   = {Intermittency and alignment in strong {RMHD} turbulence},
  journal = {Astrophys. J.},
  volume  = {807},
  pages   = {39},
  year    = {2015},
  doi     = {10.1088/0004-637X/807/1/39}
}

@article{Cruz2018,
  author  = {Cruz, F. and Alves, E. P. and Bamford, R. A. and Bingham, R. and Fonseca, R. A. and Silva, L. O.},
  title   = {Electron acceleration by wave turbulence in a magnetized plasma},
  journal = {Nat. Phys.},
  volume  = {14},
  pages   = {475--479},
  year    = {2018},
  doi     = {10.1038/s41567-018-0059-2}
}

@article{Magee2019,
  author  = {Magee, R. M. and Gota, H. and Nguyen, D. and others},
  title   = {Direct observation of ion acceleration from a beam-driven wave in a magnetic fusion experiment},
  journal = {Nat. Phys.},
  volume  = {15},
  pages   = {281--286},
  year    = {2019},
  doi     = {10.1038/s41567-018-0389-0}
}

@article{Hartouni2022,
  author  = {Hartouni, E. P. and Moore, A. S. and Crilly, A. J. and others},
  title   = {Evidence for suprathermal ion distribution in burning plasmas},
  journal = {Nat. Phys.},
  volume  = {19},
  pages   = {72--77},
  year    = {2023},
  doi     = {10.1038/s41567-022-01809-3}
}

@article{DesaiGiacalone2016,
  author  = {Desai, M. and Giacalone, J.},
  title   = {Large gradual solar energetic particle events},
  journal = {Living Rev. Sol. Phys.},
  volume  = {13},
  pages   = {3},
  year    = {2016},
  doi     = {10.1007/s41116-016-0002-5}
}

@article{Decker2005,
  author  = {Decker, R. B. and Krimigis, S. M. and Roelof, E. C. and Hill, M. E. and Armstrong, T. P. and Gloeckler, G. and Hamilton, D. C. and Lanzerotti, L. J.},
  title   = {Voyager 1 in the foreshock, termination shock, and heliosheath},
  journal = {Science},
  volume  = {309},
  pages   = {2020--2024},
  year    = {2005},
  doi     = {10.1126/science.1117569}
}

@article{Banik2024,
  author  = {Banik, U. and Bhattacharjee, A. and Sengupta, W.},
  title   = {Universal non-thermal power-law distribution functions from the self-consistent evolution of collisionless electrostatic plasmas},
  journal = {Astrophys. J.},
  volume  = {977},
  pages   = {91},
  year    = {2024},
  doi     = {10.3847/1538-4357/ad91a1}
}

@article{Zhdankin2022,
  author  = {Zhdankin, V.},
  title   = {Generalized entropy production in collisionless plasma flows and turbulence},
  journal = {Phys. Rev. X},
  volume  = {12},
  pages   = {031011},
  year    = {2022},
  doi     = {10.1103/PhysRevX.12.031011}
}

@article{Mbarek2026,
  author  = {Mbarek, R. and Gro{\v{s}}elj, D. and Philippov, A.},
  title   = {On the nonthermal power laws in magnetized turbulent plasmas},
  journal = {Astrophys. J. Lett.},
  volume  = {1003},
  pages   = {L43},
  year    = {2026},
  doi     = {10.3847/2041-8213/ae681b}
}

@article{FiskGloeckler2022,
  author  = {Fisk, L. A. and Gloeckler, G.},
  title   = {Global structure and dominant particle acceleration mechanism of the heliosheath: Definitive conclusions},
  journal = {Astrophys. J.},
  volume  = {927},
  pages   = {73},
  year    = {2022},
  doi     = {10.3847/1538-4357/ac4d2f}
}

\end{document}